# Numerical simulation of thermal properties in two-dimensional Yukawa systems

Yu. V. Khrustalyov (1 and 2), O. S. Vaulina (1)

*(1) Joint Institute for High Temperatures RAS, 125412, Izhorskaya st., 13/19, Moscow, Russia*
*(2) Moscow Institute of Physics and Technology, 141700 , Institutskiy lane, 9, Dolgoprudny, Russia*

**Abstract**. New results obtained for thermal conduction in 2D Yukawa systems. The results of numerical study of heat transfer processes for quasi equilibrium systems with parameters close to conditions in laboratory experiments with dusty plasma are presented. The Green–Kubo relations are used to calculate thermal conductivity and diffusivity coefficients. For the first time the influence of dissipation (friction) on the heat transfer in non-ideal systems is studied. New approximation is suggested for thermal diffusivity. The comparison with the existing experimental and numerical results is shown.

Получены новые данные о коэффициентах теплопереноса в двумерных (2D) системах Юкавы. Представлены результаты численного исследования теплопроводности и температуропроводности для равновесных систем с параметрами близкими к условиям лабораторных экспериментов в пылевой плазме. Для вычисления коэффициентов теплопереноса использовались формулы Грина – Кубо. Впервые исследовано влияние диссипации (трения) на процессы теплопереноса в неидеальных системах. Предложены новые аппроксимации для коэффициентов теплопроводности. Приводится сравнение полученных результатов с существующими экспериментальными, и численными данными.

## 1. Введение

Исследование транспортных свойств неидеальных систем представляет значительный фундаментальный и прикладной интерес в различных областях физики (физика плазмы, медицина, биология, физика полимеров и т.д.). [1-5]. Особый интерес вызывает исследование физических характеристик двумерных и квази- двумерных систем. (Образование таких систем, состоящих из нескольких, ~1-10, протяженных слоев заряженных пылевых частиц, часто наблюдается в условиях лабораторной пылевой плазмы ВЧ-разряда). Поскольку, помимо фундаментальных аспектов, исследования таких структур представляет особый прикладной интерес для нано- и микро-технологий, а также при разработке покрытий и материалов с заданными свойствами [3-5].

Транспортные коэффициенты являются фундаментальными константами, которые отражают природу межчастичного взаимодействия, и фазовое состояние систем. Точные теоретические предсказания транспортных коэффициентов в неидеальных жидких средах отсутствуют по настоящее время, а для анализа физических свойств таких систем обычно используют компьютерное моделирование для модельных потенциалов межчастичного взаимодействия, а также различные полуэмпирические подходы, основанные на элементарной кинетической теории, или же на аналогиях физических свойств между жидкостью и твердым телом [1-5].

Пылевая плазма представляет собой ионизованный газ, содержащий заряженные частицы вещества микронных размеров (пылевые частицы, макрочастицы). Такая плазма широко распространена в природе и образуется в ряде технологических процессов (при сгорании твердых топлив, при травлении и напылении, в производстве наночастиц и т.д.).

Макрочастицы дисперсной фазы микронных размеров (пылевые частицы), содержащиеся в плазме, могут приобретать значительный электрический заряд и формировать стационарные пылевые структуры (подобные жидкости или твердому телу) близкие к однородным трехмерным системам, или имеющие сильно неизотопный квазидвумерный характер. Благодаря своему размеру, пылевые частицы могут быть сняты видеокамерой. Это позволяет детально исследовать их динамику и делает возможным реализацию принципиально новых методов диагностики параметров пылевых частиц и окружающей их плазмы. Пылевая плазма является хорошей экспериментальной моделью для изучения различных физических процессов в диссипативных системах взаимодействующих частиц, которые представляют широкий интерес, как в области физики неидеальной плазмы, так и в других областях естественных наук. Исследования свойств пылевой плазмы могут сыграть существенную роль как в поверке существующих, так и в развитии новых эмпирических моделей в теории жидкости. Такие модели имеют огромную значимость, поскольку, в теории жидкости отсутствует малый параметр, который можно было бы использовать для аналитического описания ее физических свойств, как это возможно в случае газов. В настоящее время значительный интерес вызывают исследования физических свойств пылевой плазмы, где столкновения пылевых частиц с атомами или молекулами окружающего их нейтрального газа могут оказывать существенное влияние на транспортные характеристики комплексной среды [3-5].

С точки зрения исследования свойств пылевой плазмы особый интерес представляет экранированный кулоновский потенциал (типа Юкавы)

$$\phi = \frac{(eZ)^2}{r} \exp(-r/\lambda) \ ,$$

где $r$ – расстояние между частицами, $eZ$ – их заряд, $\lambda$ – длина экранирования. Кроме того, такая модель наиболее часто используются для моделирования отталкивания в кинетике взаимодействующих частиц (например, в медицине, биологии, физике полимеров и т.д.) [1-2].

В отличии от результатов численного моделирования процессов диффузии и вязкости в неидеальных системах Юкавы (которые широко представлены в ряде работ [6-10]), численные данные о коэффициентах теплопроводности в таких системах на настоящий момент весьма малочисленны (практически отсутствуют).

Для моделирования процессов теплопроводности в неидеальных системах частиц в настоящее время используется два основных метода молекулярной динамики. Один из них неравновесный метод (Non-Equilibrium Molecular Dynamics, NEMD), основанный на применении закона Фурье для теплового потока и включающий в себя моделирование градиентов температур. Второй равновесный подход (Equilibrium Molecular Dynamics, EMD) основан на применении соотношений Грина-Кубо в отсутствие каких-либо полей или других факторов, возмущающих систему частиц.

Отметим, что в настоящее время в научной литературе представлены расчеты теплопроводности только для трехмерных (3D) дисперсионных систем в отсутствии трения, а именно: ($i$) – для неравновесных систем Юкавы [11, 12] для случая $\kappa = l_p/\lambda < 6$, где $l_p$ – среднее межчастичное расстояние (см. Рис. 1(а), кривые $1$ и $2$); ($ii$) – для равновесных систем с другими типами (не для систем Юкавы) изотропных потенциалов [2] (см. Рис. 1(а), кривая $3$).

Следует особо подчеркнуть, что процессы теплопереноса по большей части изучены только в газообразных и кристаллических 2D и 3D системах [13]. Что же касается теплопроводности в двумерных (2D) жидкостях, то к настоящему моменту известно только две теоретические работы. Это численное моделирование тепловых свойств чисто дисперсионной системы жестких дисков (коэффициент трения для которых $\nu_{fr} = 0$) [14], и теоретическое исследование [15], где описаны трудности корректного определения коэффициентов теплопроводности для систем без трения $\nu_{fr} = 0$.

Первые результаты экспериментальных исследований тепловых свойств пылевой плазмы представлены в работах [16,17,18]. Так, в работе [16] исследовалась теплопроводность в двумерных кристаллических пылевых структурах, формирующихся в приэлектродном слое ВЧ-разряда. Измерения коэффициентов теплопроводности жидкостного пылевого монослоя в условиях близких к условиям кристаллизации плазменно-пылевой системы представлены в работе [18]. Исследования процессов теплопереноса для слабо коррелированных 3D многослойных пылевых структур описаны в работе [17]. Во всех упомянутых экспериментах наблюдалась слабая зависимость коэффициентов теплопроводности от температуры пылевых частиц. Кроме того, в работе [18] было отмечено отсутствие выраженного скачка коэффициента теплопроводности при переходе двумерного пылевого монослоя из жидкостного в кристаллическое состояние.

В настоящей работе представлены результаты первого численного исследования процессов теплопроводности в двумерных диссипативных ( $\nu_{fr} \neq 0$ ) системах Юкавы. (Еще раз отметим, что формирование таких систем обычно наблюдается в условиях лабораторной плазмы ВЧ-разряда [3, 4].)

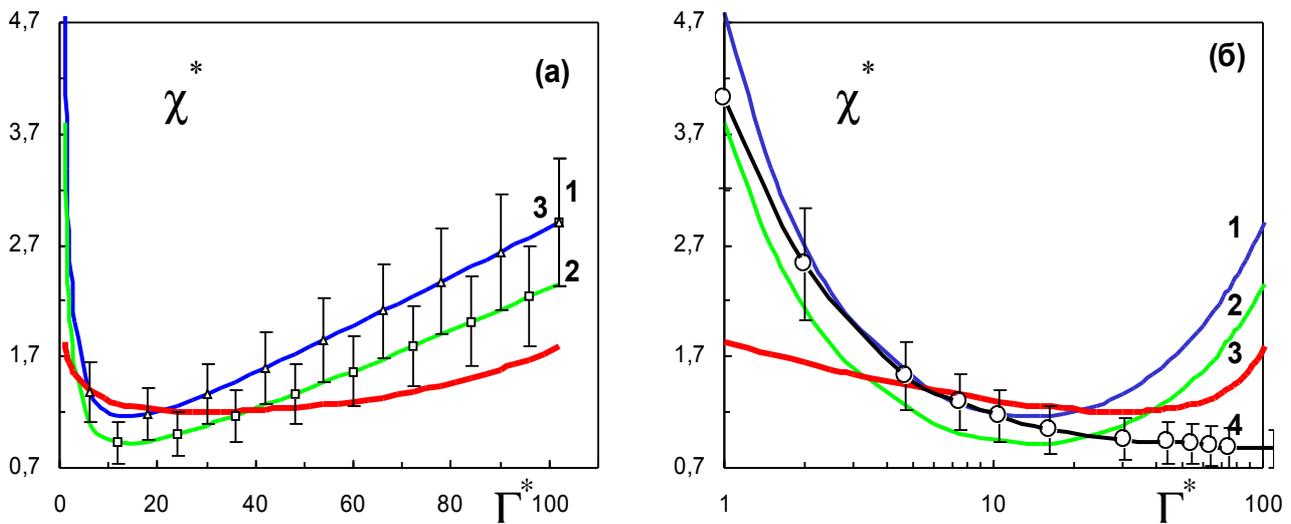

**Рис. 1.** Нормированная величина коэффициента теплопроводности $\chi^* = \chi / k_B n \omega^* l_p^2$ в зависимости от параметра $\Gamma^*$ для случаев NEMD моделирования: *1* – [7], *2* – [8] и EMD моделирования: кривая *3* – уравнение (13) для чисто дисперсионных 3D систем ( $\nu_{fr} = 0$ ), кривая *4* – данные вычислений настоящей работы для диссипативных 2D систем Юкавы в случае $\nu_{fr} \ll \omega^*$

## 2. Основные параметры

Для сравнительного анализа существующих численных данных, описывающих физические характеристики неидеальных систем, удобно ввести два безразмерных параметра, которые полностью описывают процессы массопереноса и фазовое состояние для систем с широким кругом парных изотропных потенциалов [8,9,17]. А именно, эффективный параметр неидеальности:

$$\Gamma^* = \frac{G_1 l_p^2 \phi''(l_p)}{2T}$$ (1)

и параметр масштабирования:

$$\xi = \omega^* / \nu_{fr}$$ , (2)

где

$$\omega^* = \frac{|G_2 \phi''(l_p)|^{1/2}}{(2\pi M)^{1/2}} \quad .$$

(3)

Здесь $M$ – масса пылевой частицы. $\nu_{fr}$ – коэффициент трения (т.е. эффективная частота столкновения пылевых частиц с нейтралами окружающего газа), $\phi''(l_p)$ – вторая производная парного потенциала в точке $r = l_p$, $G_1$ и $G_2$ – безразмерные константы: $G_1 = G_2 = 1$ для 3D систем и $G_1 = 1.5, G_2 = 2$ для 2D систем.

Указанные параметры отвечают за структурные, транспортные и термодинамические свойства диссипативных систем ( $\nu_{fr} \neq 0$ ) с изотропными парными потенциалами, если парный потенциал взаимодействия между частицами системы удовлетворяет условию [9]:

$$2\pi > |\phi''(l_p) \times l_p / \phi'(l_p)| \quad .$$

(4)

Здесь $\phi'(l_p)$ – значение первой производной парного потенциала в точке $r = l_p$. При этом пространственная корреляция между частицами системы не зависит от $\nu_{fr}$ и целиком определяется величиной параметра $\Gamma^*$ в диапазоне его значений от $\Gamma^* \sim 10$ до точки кристаллизации системы $\Gamma_c^*$ (т.е. до точки формирования идеального кристалла, в котором коэффициент диффузии частиц $D = 0$ ). Для двумерной неидеальной диссипативной системы, формирующих в кристаллическом состоянии примитивную треугольную решетку ($hp$), величина $\Gamma_c^* = \Gamma_{2D}^* \simeq 154 \pm 4$ [8, 9]; для трехмерных систем, образующих объемно-центрированную кубическую решетку ($bcc$): $\Gamma_c^* = \Gamma_{3D}^* \simeq 102 \pm 3$ [4, 9].

Для систем Юкавы эффективный параметр неидеальности (1) может быть представлен в форме

$$\Gamma^* = G_1 \Gamma (1 + \kappa + \kappa^2/2) \exp(-\kappa) \quad ,$$

(5)

где $\Gamma = (eZ)^2 / T l_p$ – кулоновский параметр неидеальности, $\kappa = l_p / \lambda$ – параметр экранирования. Характерную частоту (3) в этом случае можно записать в виде:

$$\omega^{*2} = G_2 \frac{(eZ)^2}{\pi l_p^3 M} (1 + \kappa + \kappa^2/2) \exp(-\kappa) \quad .$$

(6)

## 3. Существующие приближения

Коэффициенты переноса (такие, как коэффициенты диффузии $D$, теплопроводности $\chi$, вязкости $\eta$ и т.д.) характеризуют термодинамическое состояние анализируемой системы и отражают природу межчастичного взаимодействия. Для газов коэффициенты тепловой диффузии, кинематической вязкости $\nu = \eta/\rho$ и температуропроводности $\theta = \chi / \rho \, c_P$ близки по величине и могут быть записаны в виде известных аналитических соотношений (здесь $\rho = Mn$ , $n$ – концентрация атомов или молекул газа, а $c_P = C_P \, k_B / M$ – удельная теплоемкость при постоянном давлении) [1-3]. К сожалению, такие простые соотношения успешно описывают только в случае слабых взаимодействий между частицами систем, близких к идеальному газу. Наличие аналитических аппроксимаций, описывающих транспортные коэффициенты для жидкого состояния вещества, позволяет использовать известные гидродинамические модели для анализа распространения волн, сдвиговых течений, формирования вихрей и различных неустойчивостей в сильно неидеальных средах.

В случае малых отклонений исследуемой системы от состояния статистического равновесия кинетические коэффициенты линейных транспортных процессов могут быть получены из хорошо известных формул Грина-Кубо [2], которые были получены на основе теории Марковских стохастических процессов в предположении линейного отклика системы на ее слабое возмущение. В соответствие с указанным подходом, коэффициенты диффузии $D$ и теплопроводности $\chi$ в квазиравновесных системах могут быть записаны как :

$$D = \frac{1}{s} \int_0^\infty \langle \mathbf{V}(0) \mathbf{V}(t) \rangle \, dt \quad , \tag{7}$$

$$\chi = \frac{k_B \, n}{s \, T^2} \int_0^\infty \langle \delta \, \mathbf{j}_Q(0) \delta \, \mathbf{j}_Q(t) \rangle \, dt \quad . \tag{8}$$

Здесь $s$ – размерность системы, $k_B$ – постоянная Больцмана, угловые скобки обозначают усреднение по ансамблю (состоящему из $N$ частиц) и по всем интервалам времени (длительностью $t$ ), $1/s \langle \mathbf{V}(0) \mathbf{V}(t) \rangle$ – автокорреляционная функция скоростей [9], а $1/s \langle \delta \, \mathbf{j}_Q(0) \delta \, \mathbf{j}_Q(t) \rangle$ – автокорреляционная функция флуктуаций теплового потока [2], величина которого для системы, состоящей из $N$ взаимодействующих частиц может быть представлена как сумма кинетической части $\mathbf{j}_K$ и двух составляющих, которые описывают влияние внутренней $\mathbf{j}_U$ и «внешней» $\mathbf{j}_F$ потенциальной энергии (последний потенциальный член связан с собственным давлением частиц и близок по духу к рассмотрению задачи в гидродинамическом приближении) [2]:

$$\mathbf{j}_Q = \mathbf{V} \frac{1}{2} M \, \mathbf{V}^2 + \mathbf{V} \frac{1}{2} \sum_{k \in \mathrm{Sur}} \phi(r_k) - \frac{s}{4} \sum_{k \in \mathrm{Sur}} \frac{\phi'(r_k)}{r_k} (\mathbf{r}_k \, \mathbf{V}) \mathbf{r}_k \equiv \mathbf{j}_K + \mathbf{j}_U + \mathbf{j}_F \quad . \tag{9}$$

В случае изотропных парных взаимодействий флуктуации флуктуации теплового вектора $\delta\, \mathbf{j}_Q = \mathbf{j}_Q - \mathbf{j}_0$ (где $\mathbf{j}_0$ характеризует передачу тепла в статистически равновесной системе) могут быть представлены в виде:

$$\delta\, \mathbf{j}_Q = \mathbf{V}\frac{1}{2}M\left(\mathbf{V}^2 - \langle\mathbf{V}^2\rangle\right) + \mathbf{V}\frac{1}{2}\left(\sum_{k\in \text{Sur}}\phi(r_k) - \langle\sum_{k\in \text{Sur}}\phi(r_k)\rangle\right) - \frac{s}{4}\left(\sum_{k\in \text{Sur}}\frac{\phi'(r_k)}{r_k}(\mathbf{r}_k\,\mathbf{V})\mathbf{r}_k - \langle\sum_{k\in \text{Sur}}\frac{\phi'(r_k)}{r_k}(\mathbf{r}_k\,\mathbf{V})\mathbf{r}_k\rangle\right)$$ (10)

Здесь суммирование по индексу $k$ производится по всем частицам, окружающим данную частицу и влияющим на ее движение посредством парного взаимодействия. Поскольку при выводе соотношений (7) и (8) не делается предположений о характере теплового движения, они справедливы как для газов, так и для жидкостей и твердых тел. Однако в случае неидеальных сред расчет коэффициентов переноса при помощи указанных формул не допускает аналитических решений. И, как уже было упомянуто выше, для прогнозирования транспортных характеристик таких систем в настоящее время широко используются методы численного моделирования задачи и различные полуэмпирические подходы [1-5].

Численное моделирование процессов переноса в простых одноатомных жидкостях с широким кругом потенциалов взаимодействия показывает, что в случае, когда длина свободного пробега $l_{dd}$ заряженных частиц между их столкновениями сравнима со средним межчастичным расстоянием $l_p$ кинетические коэффициенты $D$ и $\chi$ можно аппроксимировать соотношениями [2]:

$$D = 0.6\, l_p\, V_T \exp(-0.8\,\sigma) \quad,$$ (11)

$$\chi = 1.5\, k_B n\, l_p\, V_T \exp(-0.5\,\sigma) \quad.$$ (12)

Здесь $V_T = \sqrt{T/M}$, а величина $\sigma$ пропорциональна конфигурационной энтропии. Значение $\sigma$ рассчитывается как сумма спектральных (частотных) характеристик системы и зависит от температуры и типа потенциала парного взаимодействия. [19]. Неоднозначность в определении величины $\sigma$ легко устранить путем совместного решения уравнений (11) и (12). Откуда для коэффициента теплопроводности можно получить следующее соотношение:

$$\chi = 1.5\, k_B n\, l_p\, V_T \left(\frac{0.6\, l_p\, V_T}{D}\right)^{5/8} \quad.$$ (13)

Нормированные значения $\chi^*(\Gamma^*) = \chi/k_B n\,\omega^* l_p^2$ в зависимости от параметра $\Gamma^*$ для чисто дисперсионных систем Юкавы ( $\nu_{fr} = 0$ ) представлены на Рис. 1 *а* и *б* в нормальном и логарифмическом масштабах, соответственно.

Кривые *1* и *2* описывают результаты неравновесных расчетов для систем с параметром экранирования $\kappa < 6$ [11, 12]. Кривая *3* – численные данные, полученные для равновесных систем по формуле (13), где для коэффициента диффузии $D$ использовалась аппроксимация, предложенная в работе [20]:

$$D^* = 1 - \frac{1}{s}\left[\frac{4\varepsilon}{1+\exp(\varepsilon)} + \frac{(s-1)}{s}\frac{\Gamma^*}{\Gamma_c^*}\right] \ , \tag{14}$$

где $\ D^* = D\left(\nu_{fr} + \omega^*\right)M/T \ $, $\ \varepsilon = 0.5 + (s-0.5)\Gamma^*/\Gamma_c^* \ $.

Нормированная величина коэффициента теплопроводности $\ \theta^* = \chi/c_P\rho\,\omega^* l_P^2 \ $ в зависимости от эффективного параметра неидеальности $\ \Gamma^* \ $ для случаев NEMD и EMD моделирования показана на Рис. 2.

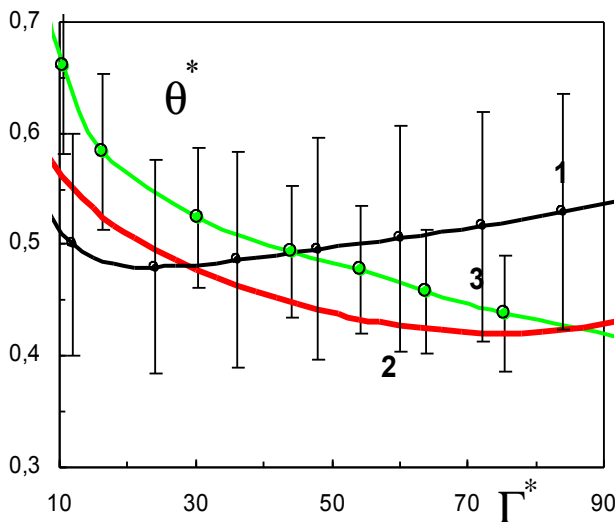

**Рис. 2** Нормированная величина коэффициента температуропроводности $\ \theta^* = \chi/c_P\,\omega^* l_P^2 \ $ в зависимости от параметра $\ \Gamma^* \ $ для случаев NEMD моделирования (кривая *1*, усредненные данные работ [7, 8]) и EMD моделирования (кривая *2*, уравнение (13)) для 3D систем при $\nu_{fr} = 0$); кривая *3* – данные вычислений настоящей работы для диссипативных 2D – систем Юкавы для $\xi \to \infty$

Здесь для вычисления коэффициента температуропроводности $\ \theta \ $ принимается во внимание, что в большинстве случаев для жидкости и твердых тел величина $\ C_P - C_V \ $ является незначительной [2, 20]. Таким образом, для оценки $\ C_P \simeq C_V \ $ можно использовать аппроксимацию для величины теплоемкости при постоянном объеме $\ C_V \ $ :

$$C_V \simeq \frac{s}{2} + \frac{2 + 4\left(\varepsilon - \frac{1}{2}\right)\dfrac{\varepsilon\exp(\varepsilon)}{1+\exp(\varepsilon)}}{1+\exp(\varepsilon)} \ , \tag{15}$$

предлагаемую в работе [20] для неидеальных систем с параметрами неидеальности в диапазоне $\ 10 < \Gamma^* < \Gamma_c^* \ $. Данная аппроксимация будет также использована ниже для анализа результатов численных исследований (см. п. 5) и для сравнения полученных данных с существующими численными и экспериментальными результатами (см. п. 6).

Дополнительно отметим, что согласно теории флуктуаций, связь между величиной $\ C_V \ $ и плотностью флуктуаций внутренней энергии системы [21] можно записать в виде следующего соотношения

$$C_V = \frac{\left\langle \left(\frac{1}{2} M V^2 + \sum_{k \in \text{Sur}} \phi_k - \left\langle \frac{1}{2} M V^2 + \sum_{k \in \text{Sur}} \phi_k \right\rangle \right)^2 \right\rangle}{N T^2} \quad . \qquad (16)$$

Обсудим разницу тепловых свойств системы, которая обнаруживается при различных методах (NEMD и EMD) моделирования (см. Рис. 1 *а* и *б* и Рис. 2). В первую очередь необходимо отметить, что в случае реальных одноатомных жидкостей в условиях близких к точке кристаллизации системы величина коэффициента теплопроводности обычно меняется в пределах 30 % – 35 % в диапазоне температур между $T_c < T < 2 T_c$ – что соответствует диапазону $50 < \Gamma^* < 100$ – см., например, данные по щелочным металлам [22]. Такие изменения находятся в соответствии с результатами EMD моделирования. Для данных NEMD, наблюдаемые изменения коэффициентов теплопроводности, по крайней мере, в два раза выше и оставляют около ~70 %.

Одной из причин существующей разницы в результатах NEMD и EMD моделирования могут являться численные ошибки, величина которых оценивается в работах [7,8] как ~ 20% (диапазон этих ошибок показан на Рис. 1 *а* и 2). Другая причина – возможное влияние градиентов концентрации частиц в случае NEMD, поскольку для механического равновесия в системе необходимо постоянство давления $P$ (а не концентрация частиц $n$). Как результат такого постоянства $P$ при наличии температурных градиентов, величина концентрации в системе может заметно меняться. В отличие от NEMD, большинство EMD вычислений (см. п. 4, п. 5) выполняются при постоянной величине средней концентрации $n$ взаимодействующих частиц.

В заключение данного параграфа отметим, что для случая $\nu_{fr} \ll \omega^*$ т.е. $\xi = \omega^* / \nu_{fr} \to \infty$ результаты MD моделирования слабодиссипативных систем и одноатомных жидкостей ( $\nu_{fr} = 0$ ) должны совпадать. Исходя из этого, возможность применения соотношения (11) для случая диссипативных систем ( $\nu_{fr} \neq 0$ ) обсуждалась в работе [20]. Однако, как будет показано ниже (см. п. 5), данное предположение оказалось ошибочным.

## 4. Параметры численной задачи

Моделирование транспорта макрочастиц в пылевой плазме требует применения метода молекулярной динамики, основанного на решении системы дифференциальных уравнений с силой Ланжевена $F_{ran}$. Для моделирования равновесных микроскопических процессов в однородных протяженных облаках взаимодействующих макрочастиц, наряду со случайными силами $F_{ran}$, которые являются источником теплового движения частиц, в системе из $N_p$ уравнений движения (где $N_p$ – количество частиц) учитывались силы парного межчастичного взаимодействия $F_{int}$:

$$M \frac{d^2 \mathbf{r}_k}{dt^2} = \sum_j F_{int}(r_{kj}) \frac{\mathbf{r}_k - \mathbf{r}_j}{r_{kj}} - M \nu_{fr} \frac{d \mathbf{r}_k}{dt} + \mathbf{F}_{ran} \quad . \qquad (17)$$

Для корректного моделирования случайных сил шаг интегрирования $\Delta t$ должен удовлетворять условию $\Delta t \ll 1 / \max \{ \nu_{fr}, \omega^* \}$. Шаг интегрирования в наших расчетах соответствовал $\Delta t = 1/20 \max \{ \nu_{fr}, \omega^* \}$, а продолжительность численных экспериментов варьировалась в диапазоне от $10 / \min \{ \nu_{fr}, \omega^* \}$ до $1000 / \min \{ \nu_{fr}, \omega^* \}$.

Расчеты были выполнены для однородных двумерных систем Юкавы с параметрами экранирования $\kappa = 1, 2, 3, 4$. Для моделирования протяженного однородного слоя задавались периодические граничные условия. Число независимых частиц в центральной ячейке $N_p$ менялось от 256 до 1225. В зависимости от числа частиц длина обрезания потенциала $l_{cut}$ менялась от $5 l_p$ до $25 l_p$. Основные расчеты были выполнены для $N_p = 256$ и $l_{cut} = 8 l_p$. Для проверки независимости результатов расчета от числа частиц и длины обрезания потенциала был выполнен анализ вычислений с различными значениями $N_p$ и $l_{cut}$. Отклонение между результатами данных расчетов не превышало численной ошибки и находилось в пределах $\pm (1\text{–}3)\%$.

Величина параметра масштабирования варьировалась от $\xi \simeq 0.1$ до $\xi \simeq 2$, типичных для условий экспериментов в плазме газовых разрядов. Величина эффективного параметра $\Gamma^*$ менялась от 0.5 до 180.

## 5. Результаты

В результате обработки численных данных были получены коэффициенты теплопроводности $\chi$ (см. уравнения (8) и (9)) как сумма кинетической $\chi_K$ двух потенциальных – $\chi_U$ и $\chi_F$ – и двух перекрестных $\chi_{FE} + \chi_{EF}$ компонент:

$$\chi = \chi_K + \chi_U + \chi_F + \chi_{FE} + \chi_{EF} \quad , \qquad (18)$$

где

$$\chi_\alpha = \frac{k_B n}{s T^2} \int_0^\infty \langle \mathbf{j}_\alpha(0) \mathbf{j}_\alpha(t) \rangle dt \quad ,$$
$$\alpha = K, U, F$$

а также

$$\chi_{FE}+\chi_{EF}=\frac{k_B n}{s\,T^2}\int_0^\infty \left\langle \left(\mathbf{j}_K(t)+\mathbf{j}_U(t)\right)\mathbf{j}_F(0)+\left(\mathbf{j}_K(0)+\mathbf{j}_U(0)\right)\mathbf{j}_F(t)\right\rangle dt \quad .$$

Нормированные автокорреляционные функции флуктуаций теплового вектора для различных параметров $\Gamma^*$ и $\xi$ показаны на Рис. 3. Нормированные функции $\chi^*(\Gamma^*)$ для различных $\xi$ показаны на Рис. 1 *б* и Рис. 4 *а*, а также приведены в Таблице 1 для случая $\xi\to\infty$ при некоторых $\Gamma^*$ совместно с величиной $(\chi_K+\chi_U+2\chi_{FE})/k_B n\omega^* l_p^2$. Величина перекрестной компоненты $\chi_{UK}+\chi_{KU}$ не оказывала существенного влияния на коэффициент $\chi$ во всем диапазоне исследуемых параметров $\xi$ и $\Gamma^*$. Величина среднеквадратичного отклонения численных данных, полученная при расчете транспортных коэффициентов в системах с различными параметрами экранирования $1<\kappa<4$, составляла менее 15%.

Анализ полученных данных показал, что величина коэффициента теплопроводности полностью определяется двумя безразмерными параметрами $\xi$ и $\Gamma^*$. Для всех рассмотренных случаев, перенос тепла за счет «внешней» потенциальной энергии $\chi_F$ (связанной с собственным давлением частиц в системе) не зависел от коэффициента трения $\nu_{fr}$ и оставался практически постоянным для всех $\Gamma^*>5$, а отношение

$$\frac{\chi_F}{k_B n\omega^* l_p^2}\simeq 0.75\pm 0.12$$

не зависело от температуры частиц в системе.

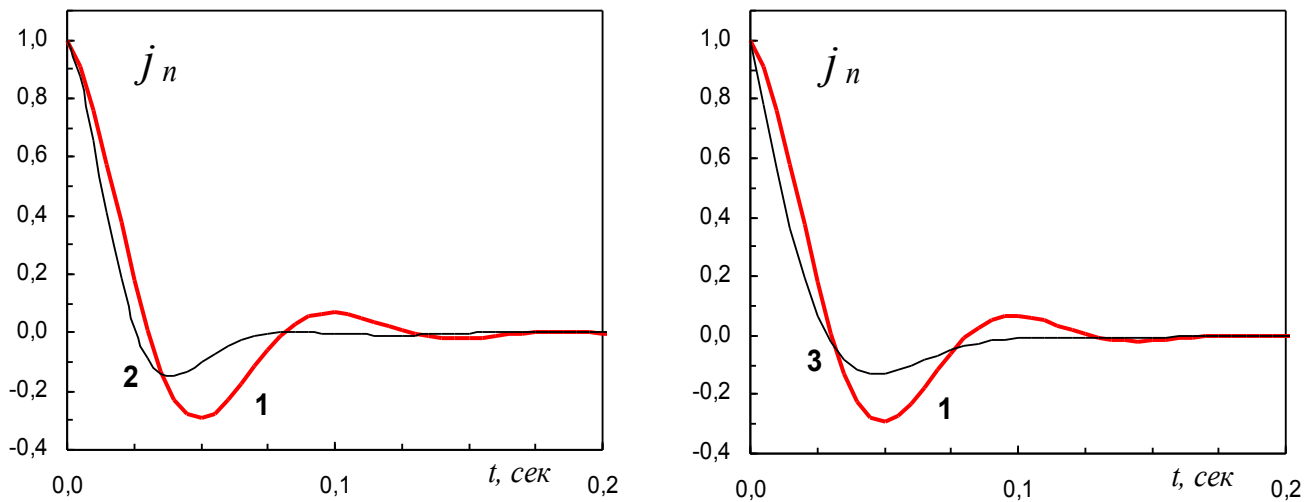

**Рис. 3.** Нормированные автокорреляционные функции флуктуаций теплового вектора $j_n(t)=\langle \mathbf{j}_Q(0)\mathbf{j}_Q(t)\rangle/\langle \mathbf{j}_Q(0)\mathbf{j}_Q(0)\rangle$ для: *1 –* $\xi=1$ , $\Gamma^*=75$ ; *2 –* $\xi=1$ , $\Gamma^*=10$ ; *3 –* $\xi=0.25$ , $\Gamma^*=75$

| $\Gamma^*$ | 1 | 4.7 | 10.5 | 16.2 | 30.4 | 54.2 | 75.3 | 111 | 160 |
|---|---|---|---|---|---|---|---|---|---|
| $\dfrac{\chi_K + \chi_U + 2\chi_{FE}}{k_B n \omega^* l_p^2}$ | 3.26 | 0.78 | 0.42 | 0.29 | 0.204 | 0.162 | 0.132 | 0.119 | 0.110 |
| $\dfrac{\chi}{k_B n \omega^* l_p^2}$ | 4.01 | 1.53 | 1.17 | 1.04 | 0.96 | 0.91 | 0.88 | 0.87 | 0.86 |

**Таблица 1.** Нормированные коэффициенты теплопроводности в зависимости от $\Gamma^*$ для чисто дисперсионных ($\xi \to \infty$) двумерных систем Юкавы

Теплоперенос внутренней энергии (а именно, ее кинетической $\chi_K$ потенциальной $\chi_U$ и перекрестной $2\chi_{FE}$ составляющих) был обратно пропорционален сумме характерных частот $\nu_{dis} = \nu_{fr} + \omega^*$, при этом величина

$$\chi_U^* = \chi_U \frac{1 + \xi^{-1}}{k_B n \omega^* l_p^2} \simeq 0.110 \pm 0.015$$

сохранялась постоянной и слабо зависела от параметра неидеальности при $\Gamma^* > 5$. Таким образом, можно полагать, что нормированная величина коэффициента теплопроводности $\chi^*$ практически полностью определяется значением эффективного параметра неидеальности и параметра масштабирования $\Gamma^* > 5$ и может быть аппроксимирована как

$$\chi^* \simeq \frac{A}{1 + \xi^{-1}} + \frac{0.11}{1 + \xi^{-1}} + 0.75 \quad , \tag{19}$$

где $A$ — коэффициент, величина которого определяется значением суммы $\chi_K + \chi_U + 2\chi_{FE}$ для случая чисто дисперсионных систем $\xi \to \infty$, а именно (см. таблицу 1):

$$A = \frac{\chi_K + \chi_U + 2\chi_{FE}}{k_B n \omega^* l_p^2} \quad .$$

Принимая во внимание, что $\chi_K / k_B \rho \propto C_P D \simeq C_V D$ для $\Gamma^* > 5$, и используя процедуру подгонки численных данных, можно найти более удобную аппроксимацию для коэффициента теплопроводности в двумерных неидеальных системах, не требующую дополнительной информации для чисто дисперсионного случая (т.е. о коэффициенте $A$).

$$\chi^* \simeq \frac{\pi C_V D^*}{\Gamma^* (1 + \xi^{-1})} + \frac{0.11}{1 + \xi^{-1}} + 0.75 \quad . \tag{20}$$

Сравнение соотношения (20) с численными данными для различных параметров масштабирования $\xi$ показано на Рис. 4. Легко увидеть хорошее соответствие предлагаемой аппроксимации результатам численного расчета коэффициентов теплопроводности методом EMD моделирования.

В заключение, обсудим два дополнительных (не менее важных) результата проведенных исследований, которые могут быть полезны для развития новых методов диагностики пылевой компоненты плазмы, включая: выбор условий для корректного определения коэффициентов теплоемкости и теплопроводности по экспериментальным данным о движении частиц, а также для разработку новых методик восстановления упомянутых коэффициентов для слабо-коррелированных неидеальных систем без априорной информации о форме потенциала межчастичного взаимодействия.

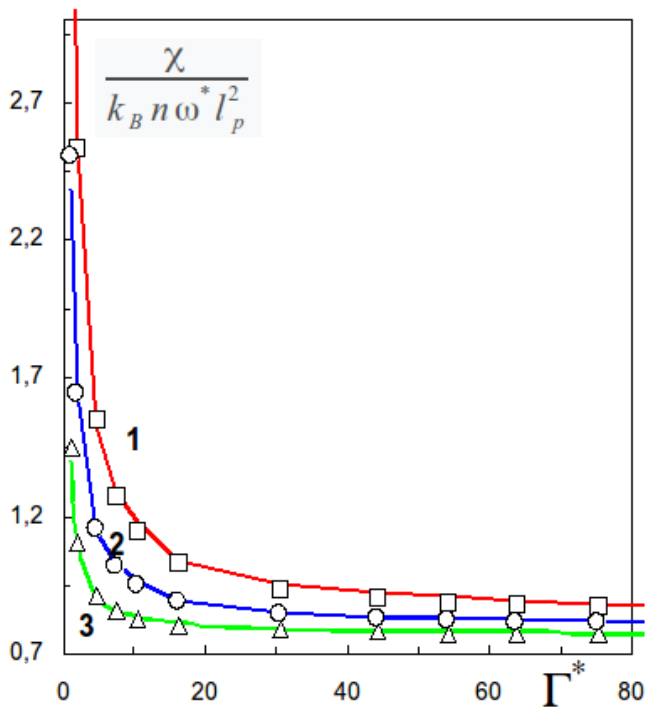

**Рис. 4.** Нормированная величина коэффициента теплопроводности $\chi^*$ в зависимости от $\Gamma^*$ для диссипативных 2D систем Юкавы с различными параметрами масштабирования $\xi$ : *1 –* $\xi \to \infty$ *, 2 –* $\xi = 1$ *, 3 –* $\xi = 0.25$ *; символы – результаты численных расчетов, а линии – аппроксимация* (20)

Первый результат заключается в исследовании связи между кинетической частью $\chi_K$ коэффициента теплопроводности и коэффициентом диффузии $D$ частиц. Так, было получено, что с уменьшением эффективного параметра неидеальности $\Gamma^* < 2$ отношение кинетической части нормированного коэффициента теплопроводности $\chi_K / k_B n$ к коэффициенту диффузии $D$ в исследуемых системах стремиться к величине близкой к коэффициенту теплоемкости идеальной системы при постоянном давлении (см. Рис. 5):

$$\frac{\chi_K}{k_B n D} \to C_P \simeq 2 \quad .$$

При этом с уменьшением параметра $\Gamma^*$ величина кинетической части $\chi_K$ коэффициента теплопроводности приближается к его истинному (полному) значению $\chi$ : $\chi_K \to \chi$ , что позволяет проводить простое экспериментальное определение данного коэффициента при $\Gamma^* < 2$ на основе прямых измерений скоростей частиц, не опираясь на какую-либо информацию о форме потенциала их межчастичного взаимодействия.

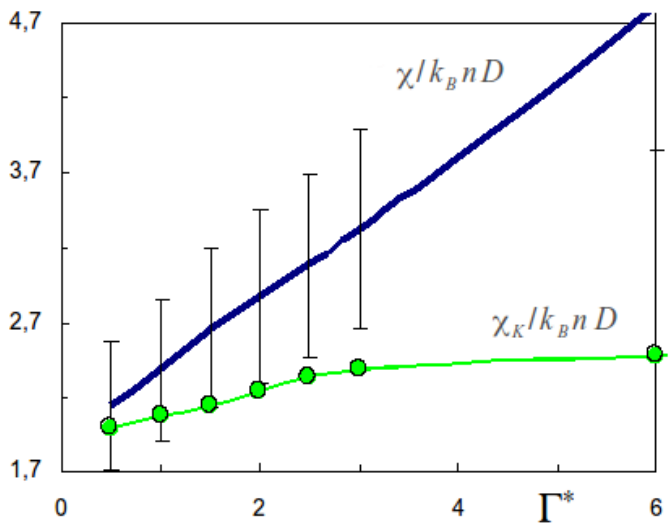

**Рис. 5.** Отношения $\chi/k_B n D$ и $\chi_K/k_B n D$ в зависимости от параметра $\Gamma^*$.

Второй результат – это численные данные о коэффициенте теплоемкости $C_V$ неидеальных систем, полученные при помощи метода флуктуаций (16). Сравнения результатов вычисления коэффициентов теплоемкости $C_V$, полученных различными методами моделирования, представлено на Рис. 6. Там же показано сравнение величины $C_V$ с разницей величиной $\chi_K/k_B n D - 1$. Можно легко заметить, что ранее предложенная аппроксимация (15) хорошо описывает расчеты теплоемкости системы с параметрами $\Gamma^* > 10$, а также то, что для слабо-коррелированных систем величина

$$C_V \simeq \frac{\chi_K}{k_B n D} - 1$$

с параметрами $\Gamma^* < 20$. Последнее обстоятельство позволяет определять величину коэффициента теплоемкости в таких системах, не опираясь на предварительную информацию о потенциале межчастичного взаимодействия.

**6. Сравнение с существующими численными и экспериментальными данными**

Сравнение результатов наших расчетов коэффициентов теплопереноса (теплопроводности и температуропроводности) в двумерных диссипативных системах Юкавы с существующими численными данными представлены на Рис. 1 *б* и 2. Возможные причины отличия между равновесными и неравновесными методами моделирования задачи подробно рассматриваются в п. 3 настоящей работы.

Здесь мы остановимся на сравнении наших результатов с результатами моделирования трехмерных равновесных систем. В данном случае отсутствие выраженного минимума функции $\chi^*(\Gamma^*)$ для 2D систем (в отличие от результатов, полученных для 3D систем, см. Рис 1 *б*) легко объяснить, если учесть, что в системах различной размерности только коэффициенты температуропроводности $\theta$ (а не коэффициенты теплопроводности $\chi$) должны иметь близкие температурные зависимости [2]. Такое предположение хорошо подтверждается представленными результатами моделирования двумерной задачи. Легко увидеть (см. Рис. 3 кривые *2, 3*), что поведение $\theta^*(\Gamma^*)$ является близким для равновесных 2D и 3D систем, и практически одинаковым с точностью до погрешности численного моделирования нашей задачи.

Сравнение наших численных расчетов коэффициента $\theta^*(\Gamma^*)$ (20) и расчетов по формуле, предлагаемой в теории молекулярных жидкостей (13) с экспериментальными данными работы [17] представлено на Рис. 7 для параметров упомянутых экспериментов: $l_p = 450 \pm 50\,\textit{мкм}$, $\nu_{fr} \simeq 30\,c^{-1}$. Здесь мы приняли во внимание различную нормировку коэффициентов для систем с разной размерностью $\omega^* = 7.7 \pm 0.8\,c^{-1}$, $\xi = 0.25$ для 3D систем и $\omega^* = 10.9 \pm 1.1\,c^{-1}$, $\xi = 0.35$ для 2D структур (см. п. 2). Легко увидеть, что предлагаемая аппроксимация (20) находится в хорошем согласии с рассматриваемыми экспериментальными данными, в отличие от соотношения (13), полученного из теории молекулярных жидкостей.

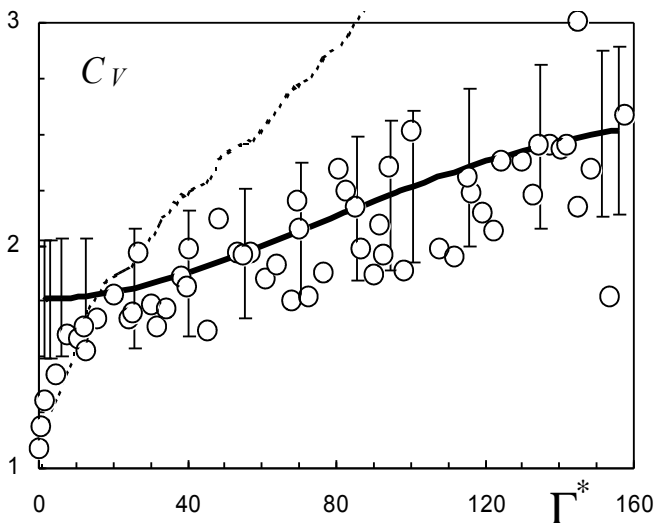

**Рис. 6.** Коэффициенты теплоемкости $C_V$, полученные путем численных расчетов: сплошная линия – аппроксимация (15), ○ – из теории флуктуаций (16) и пунктирная линия – $\chi_K / k_B n\,D - 1$

Постоянство коэффициента теплопроводности в сильно коррелированных двумерных жидкостных пылевых системах вблизи их точки кристаллизации было найдено в экспериментах в приэлектродной плазме ВЧ-разряда [18]. Чтобы объяснить этот полученный

«нетривиальный» результат авторы работы предположили, что главную роль в условиях наблюдаемых экспериментов играло рассеяние фононов (например, вследствие динамической неоднородности системы[23]). Однако представленные здесь результаты (см. Рис. 1 *б* и Рис. 3) позволяют предположить другую причину для наблюдаемого явления, которая заключается во влиянии размерности системы (исследуемого пылевого монослоя) на температурную зависимость коэффициента теплопроводности. Следует отметить, что для параметров упомянутых экспериментов – $l_p \simeq 660\,мкм$ , $\omega^* \simeq 17\,c^{-1}$ , $\nu_{fr} \simeq 0.87\,c^{-1}$ и $\Gamma^* \sim 100..150$ [18] – величина $c_P \theta = \chi/\rho$ , рассчитанная из аппроксимации (20) находится в диапазоне от 6.4 до 7.0 $мм^2/c$, что хорошо согласуется с результатами ее экспериментального определения $c_P \theta \simeq 9\,мм^2/c$ [18].

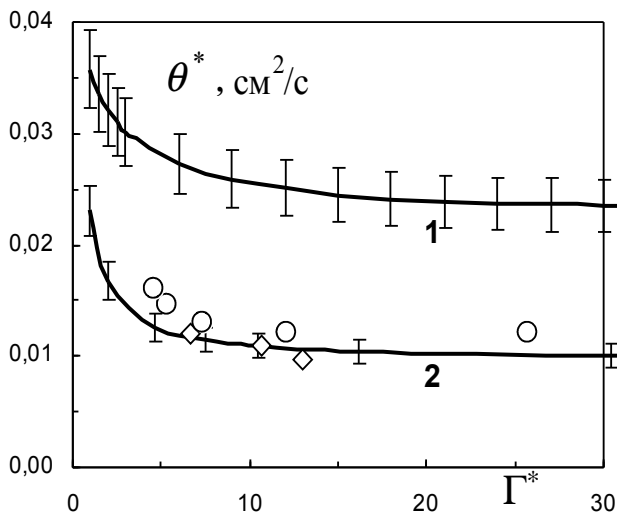

**Рис. 7.** Функция $\theta^*(\Gamma^*)$ : *символы* – для экспериментов [17], а также численные данные для условий экспериментов *сплошные линии*:
*1* – уравнение (13), (3D: $\omega^* = 7.7 \pm 0.8\,c^{-1}$ , $\xi = 0.25$ )
*2* – уравнение (20), (2D: $\omega^* = 10.9 \pm 1.1\,c^{-1}$ , $\xi = 0.35$ )

## 7. Заключение

Получены новые данные о коэффициентах теплопереноса (теплопроводности и температуропроводности) в двумерных диссипативных системах Юкавы. Вычисления были выполнены на основе соотношений Грина – Кубо для равновесных систем с параметрами близкими к условиям лабораторных экспериментов в пылевой плазме. Влияние диссипации (трения) на процессы теплопереноса в неидеальных системах исследовалось впервые. Было получено, что величина коэффициентов теплопереноса полностью определяется двумя безразмерными параметрами $\Gamma^*$ и $\xi$ . Предложены новые аппроксимации для коэффициентов теплопроводности. Представленные результаты находятся в хорошем соответствии с результатами численного моделирования равновесных трехмерных систем без диссипации, а также с существующими экспериментальными данными.

Было найдено, что с уменьшением эффективного параметра неидеальности $\Gamma^* < 2$ кинетическая часть коэффициента теплопроводности $\chi_K$ приближается к его истинному (полному) значению $\chi$, а величина коэффициента теплоемкости $C_V$ стремится к величине $\chi_K / k_B n D - 1$ для $\Gamma^* < 20$. Это позволяет проводить простое экспериментальное определение упомянутых коэффициентов для слабо коррелированных систем на основе прямых кинетических измерений скоростей частиц, не опираясь на предварительную информацию о величине их межчастичного взаимодействия.

Можно предположить, что результаты представленного исследования могут быть обобщены на двумерные систем с широким кругом изотропных парных потенциалов, которые удовлетворяют условию (4). Представленные результаты могут быть также полезны для разработки новых методов пассивной диагностики и развития существующих экспериментальных техник исследования разнообразных неидеальных систем таких, как, например, пылевая плазма, биологические и медицинские растворы, полимеры и другие коллоидные системы.